\documentclass[a4paper, 10pt, conference]{IEEEtran}
\IEEEoverridecommandlockouts
\usepackage{cite}
\usepackage{amsmath,amssymb,amsfonts}
\usepackage{algorithmic}
\usepackage{graphicx}
\usepackage{textcomp}
\usepackage{xcolor}
\usepackage{comment}
\usepackage{tikz}

\def\BibTeX{{\rm B\kern-.05em{\sc i\kern-.025em b}\kern-.08em
    T\kern-.1667em\lower.7ex\hbox{E}\kern-.125emX}}
\begin{document}

\title{Investigating the impact of virtual element misalignment in collaborative Augmented Reality experiences}

\author{
 \IEEEauthorblockN{Francesco Vona$^{1}$, Michael Stern$^1$, Navid Ashrafi$^{1}$, 
 Tanja Kojić$^{2}$, \\
 Sina Hinzmann$^1$, David Grieshammer$^{1}$, Jan-Niklas Voigt-Antons$^{1}$\\}
 \IEEEauthorblockA{$^1$Immersive Reality Lab, Hamm-Lippstadt University of Applied Sciences, Hamm, Germany\\
$^2$Quality and Usability Lab, Technische Universität Berlin, Berlin, Germany
}
}


\maketitle

\begin{abstract}
The collaboration in co-located shared environments has sparked an increased interest in immersive technologies, including Augmented Reality (AR). Since research in this field has primarily focused on individual user experiences in AR, the collaborative aspects within shared AR spaces remain less explored, and fewer studies can provide guidelines for designing this type of experience. This article investigates how the user experience in a collaborative shared AR space is affected by divergent perceptions of virtual objects and the effects of positional synchrony and avatars. For this purpose, we developed an AR app and used two distinct experimental conditions to study the influencing factors. Forty-eight participants, organized into 24 pairs, participated in the experiment and jointly interacted with shared virtual objects. Results indicate that divergent perceptions of virtual objects did not directly influence communication and collaboration dynamics. Conversely, positional synchrony emerged as a critical factor, significantly enhancing the quality of the collaborative experience.
On the contrary, while not negligible, avatars played a relatively less pronounced role in influencing these dynamics. These findings can potentially offer valuable practical insights, guiding the development of future collaborative AR/VR environments.
\end{abstract}
\newcommand\copyrighttext{%
  \footnotesize \textcopyright 2024 IEEE. Personal use of this material is permitted. Permission from IEEE must be obtained for all other uses, in any current or future media, including reprinting/republishing this material for advertising or promotional purposes, creating new collective works, for resale or redistribution to servers or lists, or reuse of any copyrighted component of this work in other works. DOI and link to original publication will be added as soon as they are available.}
  
\newcommand\copyrightnotice{%
\begin{tikzpicture}[remember picture,overlay,shift={(current page.south)}]
  \node[anchor=south,yshift=10pt] at (0,0) {\fbox{\parbox{\dimexpr\textwidth-\fboxsep-\fboxrule\relax}{\copyrighttext}}};
\end{tikzpicture}%
}

\copyrightnotice

\section{Introduction}
According to Azuma\cite{Azuma1997-wd}, Augmented Reality (AR) is defined as a technology with three characteristics: 1) it combines real and virtual, 2) it is interactive in real-time, and 3) registration of digital with real objects.

These features not only enhance individual user experience but implicitly create a framework conducive to collaboration. This framework allows users, whether in the same physical space or remotely, to interact with a mutually understood and shared augmented context, thus promoting a cohesive and synchronized working environment.

Consequently, the concept of shared perception is fundamental to AR \cite{Billinghurst2022}. This involves the mutual understanding of virtual content among users. For example, when multiple individuals observe a virtual object placed on an actual table, it fosters an environment where they can collectively discuss, adjust, and interact with that object \cite{Billinghurst2022}.

Collaborative Augmented Reality (AR) presents numerous challenges and considerations for developers and users. One central concern is synchronization, where maintaining consistent alignment of virtual content across all user devices is critical to the collective experience \cite{Youjeong2012}. Any lag or misalignment can significantly disrupt the collaborative interaction. Additionally, heightened spatial awareness is essential, especially in co-located AR scenarios, as participants must be cognizant of both virtual and real-world elements, as well as the actions of fellow participants, which may lead to cognitive strain in immersive environments \cite{bowman2007}. 

Beyond technical and cognitive challenges, interpersonal dynamics are integral to collaborative AR experiences. Unlike single-user AR scenarios, collaborative AR involves complex interactions among multiple users, each contributing unique perspectives and reactions to the virtual environment, unpredictably influencing the overall experience and collaborative outcomes \cite{Billinghurst2022}.

Inconsistencies in collaborative AR experiences pose a significant challenge, as revealed in various studies, ranging from technical glitches to bad user feedback \cite{7836453, 10.3389/frvir.2021.578080, mti6090075}.
This study seeks to address this gap by investigating the impact of virtual element misalignment in collaborative AR experiences. The research goals guiding this work are formulated as follows:

\textbf{RG1. Influence of virtual element misalignment on user communication and collaboration:} This study assesses the impact of virtual object misalignment on the communication dynamics among users in a shared AR setting. We explore how variations in the perception of object positions affect collaborative tasks and user interactions.

\textbf{RG2. Influence of virtual element and avatar misalignment on user communication and collaboration:} We investigate the function of avatars within an AR environment, focusing on their ability to guide user orientation and teamwork. Specifically, the study examines whether avatars can compensate for misalignments, reducing the necessity for precise positional synchrony during collaborative tasks.

To address these questions, we introduced two experimental settings. The first involves introducing virtual element misalignment in a cooperative task requiring sorting virtual objects by color and shape. In the second setting, the misalignment of virtual objects and avatars is gradually introduced in a different cooperative task.

\section{Theoretical Background}
Collaboration within shared immersive environments accentuates these mediums' distinctive advantages and challenges. A body of research examining these dynamics and exploring alternative collaborative approaches \cite{Weng2021-ae, Pidel2020CollaborationIV, Billinghurst2022} contributes significantly to understanding this domain. Billinghurst et al. \cite{Billinghurst2022} delineate the evolution of collaborative Augmented Reality (AR) since the early 2000s, noting initial solutions akin to traditional videoconferencing systems. For instance, Schmalstieg et al. \cite{Szalavari-1998-Stu} present the "Studierstube" architecture, catering to multi-user AR for visualization, presentation, and educational purposes. This system facilitates collaborative work by offering synchronized 3D stereoscopic graphics via lightweight see-through head-mounted displays, allowing users to adjust viewpoints and data layers independently. Such features strengthen cooperative efforts, particularly among visualization experts.

Other systematic investigations, such as that by Pidel et al. \cite{Pidel2020CollaborationIV}, offer comprehensive reviews of collaborative efforts in virtual and augmented reality, elucidating specific benefits and challenges. This research sheds light on often overlooked forms of collaboration, including asynchronous methods and joint efforts integrating AR and VR. While synchronous co-located and remote AR collaborations have received ample attention, asynchronous methods remain relatively unexplored. These methods unfold over time, removing the requirement for simultaneous participation, and often involve activities such as annotating and reviewing content post-engagement \cite{7836453}.

The term "co-presence" denotes users' shared local physical presence in both the real world and the shared AR environment, serving as a pivotal concept in investigating collaboration in AR settings. Efforts have been made to develop assessment tools for co-presence, aiming to integrate virtual elements seamlessly into users' physical surroundings \cite{regenbrecht2021measuring}. Assessing this psychological "presence" becomes vital, especially in shared virtual environments, with studies highlighting the role of Mixed Reality (MR) in shaping collaborative settings \cite{inbook}. Moreover, the research underscores the significance of social presence in cooperative AR, emphasizing design insights to enhance cooperation support \cite{osmers2021}. Different interaction techniques such as hand-tracking \cite{voigt2020influence} might also impact cooperative AR experiences, but this potential influence is still understudied.

Avatars are commonly employed in AR/VR environments to represent other users' positions and gestures within the shared space. Studies have investigated their influence on collaboration and social interaction \cite{shisha2021}, as well as optimal visual representation in augmented reality \cite{8797719}. The latter study explores avatar aesthetics' impact on Social Presence and user perceptions in AR telepresence scenarios, revealing preferences for realistic full-body avatars in remote collaborations.
These theoretical underpinnings set the stage for subsequent experiments and analyses to expand our understanding of collaboration in AR environments. Notably, there is a shortage of research investigating the effects of virtual element misalignment on user experiences in AR despite their potential relevance for designers and developers. Hence, our study aims to fill this gap, recognizing the significance of such exploration for advancing AR design practices.

\section{Methods}
\subsection{Participants}
Forty-eight (n = 48) participants were recruited voluntarily, of which 27 (56.25\%) identified themselves as male and 21 (43.75\%) as female. The subjects' age ranged from 21 to 51 years, with an average age of 29.00 (\textit{SD} = 7.19) years. Concerning previous experience in using VR/AR systems, 22 (45.83\%) of them had no experience, 15 (31.25\%) had limited experience, 6 (12.5\%) had some experience, and 5 (10.42\%) had extensive experience.

\subsection{Experiment Setup}
\begin{figure}[tb]
 \centering 
 \includegraphics[width=\columnwidth]{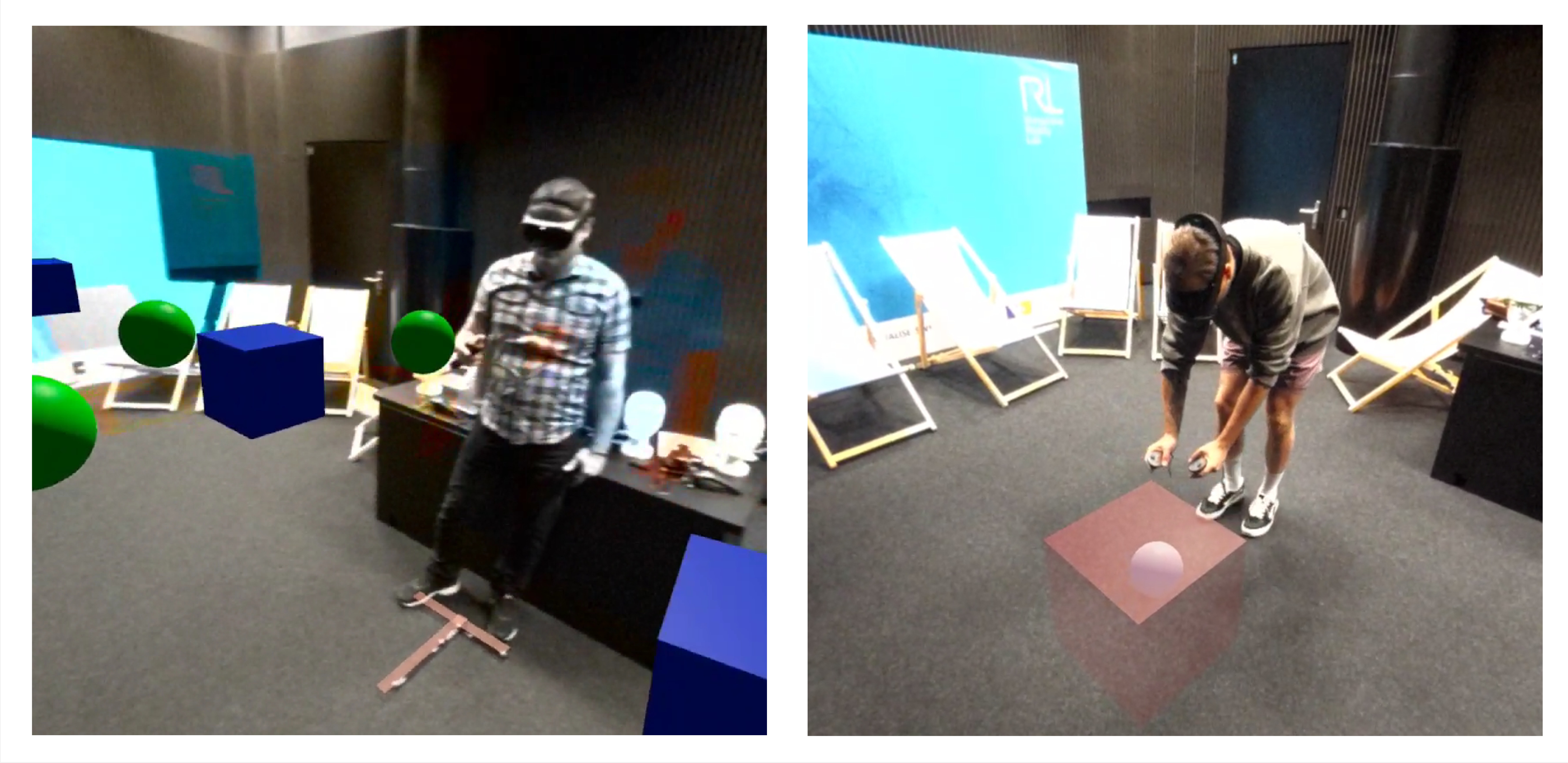}
\caption{The two tasks for the experiment setups. On the left, the task executed in the first experiment setup, in which the users have to sort blue cubes and green spheres by color and shape. On the right, the task executed in the second experiment, where the user verbally guides their partner to place a sphere into a semi-transparent cube visible only to the guiding subject.}
 \label{fig:task}
\end{figure}

\begin{figure}[tb]
 \centering 
 \includegraphics[width=\columnwidth]{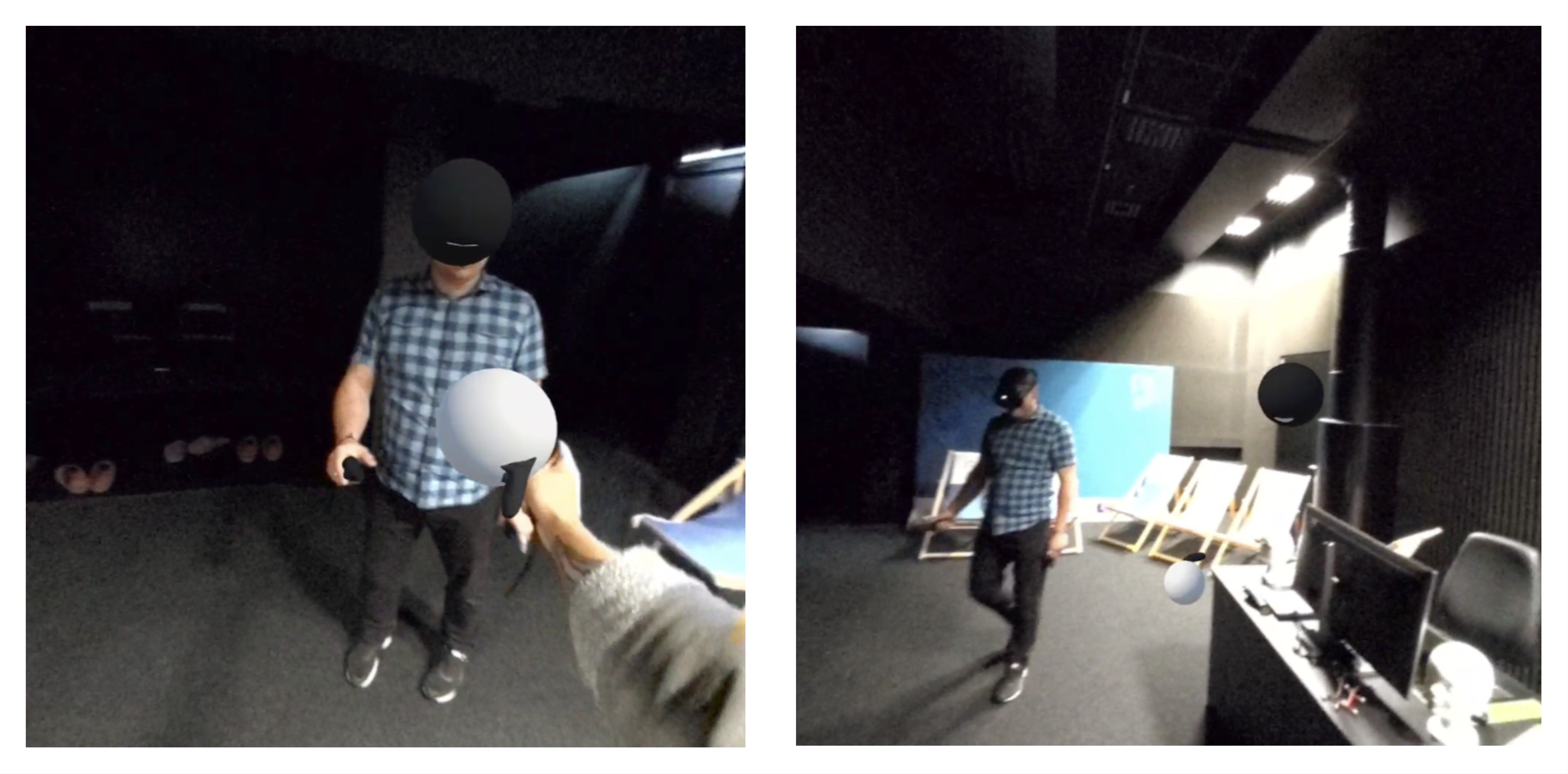}
\caption{The two conditions of the second setup with avatars. On the left, avatars are aligned with the user (``Sync-w-A"), on the right, avatars are not aligned (``ASync-w-A").}
 \label{fig:avatar}
\end{figure}

Two different experimental comparisons were designed for the study, which the users completed in a specific order in groups of two. The respective tasks in these comparisons resembled a game to provide the users with an enjoyable experience. Each comparison aims to answer one of the research goals highlighted before. The study was approved by the local ethics commission. \\
\textbf{Comparison 1: RG1.Influence of virtual element misalignment on user communication and collaboration.} The first part of the study aimed to observe participants in a cooperative task involving sorting virtual objects (green balls and blue cubes) by color and shape (Figure\ref{fig:task} left). In the \textit{``Convergence"} condition, users sorted identical objects, serving as a baseline. In the \textit{``Divergence"} condition, one object differed in appearance, potentially causing uncertainty and altering communication and cross-checking strategies. 
\\
\textbf{Comparison 2: RG2.Influence of virtual element and avatar mis-
alignment on user communication and collaboration}
The second comparison investigated four different conditions to explore the influence of positional synchrony and avatars on collaborative augmented reality (AR) tasks. Participants were instructed to verbally guide their partner to place a sphere into a semi-transparent cube visible only to the guiding subject (Figure\ref{fig:task} right). The four conditions were:
\begin{enumerate}
    \item ``Sync-w/o-A": Users' positions were synchronized, and avatars were deactivated.
    \item ``Sync-w-A": Users' positions were synchronized, and avatars were activated and synchronized (Figure\ref{fig:avatar} left).
    \item ``ASync-w/o-A": Users' positions were asynchronized, and avatars were deactivated.
    \item ``ASync-w-A": Users' positions were asynchronized, and avatars were activated and asynchronized (Figure\ref{fig:avatar} right).
\end{enumerate}
\subsection{Questionnaires}
To collect data on the general user experience in terms of learnability and comprehensibility of the system, the evaluation of physical activity during task performance, the perception of presence in the virtual environment, and the general technical affinity of the subject, a set of questionnaires were selected that are listed below. These standardized questionnaires were not used entirely, but only some questions were selected
\begin{itemize}
    \item \textbf{Demographic data and previous experience with VR systems.} Questions about demographic data, including age and gender, and self-assessment of experience with VR/AR were collected before the start of the experiments.
    \item \textbf{Affinity for Technology Interaction Scale (ATI).} The ATI scale assesses a person's tendency to engage in intensive technology interaction or avoid it actively \cite{ati2017}.
    \item \textbf{User Experience Questionnaire (UEQ),} The UEQ questions capture a comprehensive impression of the user experience and measure both classic usability aspects and user experience aspects. From the UEQ, all the items from the dimension \textit{``Perspicuity"} and another item from the dimension \textit{``Dependability"} were selected since \textit{``Perspicuity"} shows how easy people understand the product and \textit{``Dependability"} gives an idea about it seeming trustworthy \cite{ueq2008}. 
    \item \textbf{Flow State Scale (FSS)}. The FSS measures the degree of a person's so-called "flow state." The flow state is a psychological state that a person experiences while fully immersed in physical activity \cite{Csikszentmihalyi_undated-fs, Csikszentmihalyi1992-fh}. It comprises 36 items from 9 dimensions \cite{Jackson1996-yj}. From the FSS one item from each dimension was taken, namely Challenge-skill balance, Action-awareness merging, Unambiguous feedback, Sense of control, and Autotelic experience. Only \textit{``Transformation of Time"} and \textit{``Loss of Self-Consciousness"} were discarded since we were not interested in the time spent during the experience. We could get better data on the ``Loss of Self-Consciousness" through the IPQ, whereas for ``Concentration on Task" and for ``Clear Goals" respectively (2 and 3 questions selected).
    \item \textbf{iGroup Presence Questionnaire (IPQ).} The IPQ is a scale for measuring the sense of presence in a virtual environment (VE) \cite{ipq2003}. Only one item was adapted for this test environment, changing the ``computer-generated world" to ``augmented reality" to refer to the comparisons. 
    \item \textbf{Additional questions.} In addition, three further questions with free-response options were asked to capture subjective experiences and thoughts used for the qualitative analysis. Moreover, the experimenter took written comments and notes during the study execution.
    \begin{enumerate}
        \item ``How has communication changed for you when you and others in the same room did not see the same objects?"
        \item ``Assuming that we mostly have digital glasses and contact lenses instead of smartphones in the future, how do you think your communication changes when you and other people can see different things while interacting in the same room?"
        \item ``If we could do anything related to Augmented Reality without technical limits, and we would develop it specifically for you, what would an ideal system look like for you?"
    \end{enumerate}
\end{itemize}

\subsection{Procedure}
The experiments took place in a controlled laboratory environment at the university. Participants arrived in pairs at the designated time, where an on-site staff member was responsible for documentation, observation, addressing potential questions, and preparing the experimental conditions.
Upon arrival, participants signed the consent form and completed the initial section of the questionnaire (Demographic data and previous experience with VR systems plus the ATI scale). A brief introduction to the hardware's functionality was provided, followed by an explanation of the synchronization process conducted immediately before each experiment. Synchronization involved establishing positional synchronicity by aligning virtual position markers with a real-world floor marker.
For conditions such as \textit{Convergence}, \textit{Divergence}, \textit{Synch-w-A}, and \textit{Synch-w/o-A}, participants followed instructions to achieve synchronicity. Random asynchrony was introduced based on staff-provided orientations in cases of \textit{ASynch-w-A} and \textit{ASynch-w/o-A}. 
Following the completion of the initial phase, participants were handed printed instructions outlining the tasks for the two conditions of the first experiment. The sequence commenced with the \textit{``Convergence"} condition, where participants engaged in the specified tasks collaboratively. Subsequently, the experiment progressed to the \textit{``Divergence"} condition, with participants navigating through the designated tasks in the given order.
After completing the first set of experimental conditions, participants were presented with printed instructions for the subsequent phase. This time, instructions were provided for the four conditions of the second experiment. Participants systematically proceeded to execute the \textit{``Synch-w/o-A"}, \textit{``Synch-w-A"}, \textit{``ASynch-w/o-A"}, and \textit{``ASynch-w-A"} conditions in the specified sequence.
After each experiment, participants filled out the relevant sections of the questionnaires, including the FSS and UEQ questions.
After completing all experiments, participants filled out the concluding section of the questionnaire, which included the IPQ and the three open-ended questions regarding their subjective perception.

\section{Results}

\begin{figure}[tb]
 \centering 
 \includegraphics[width=\columnwidth]{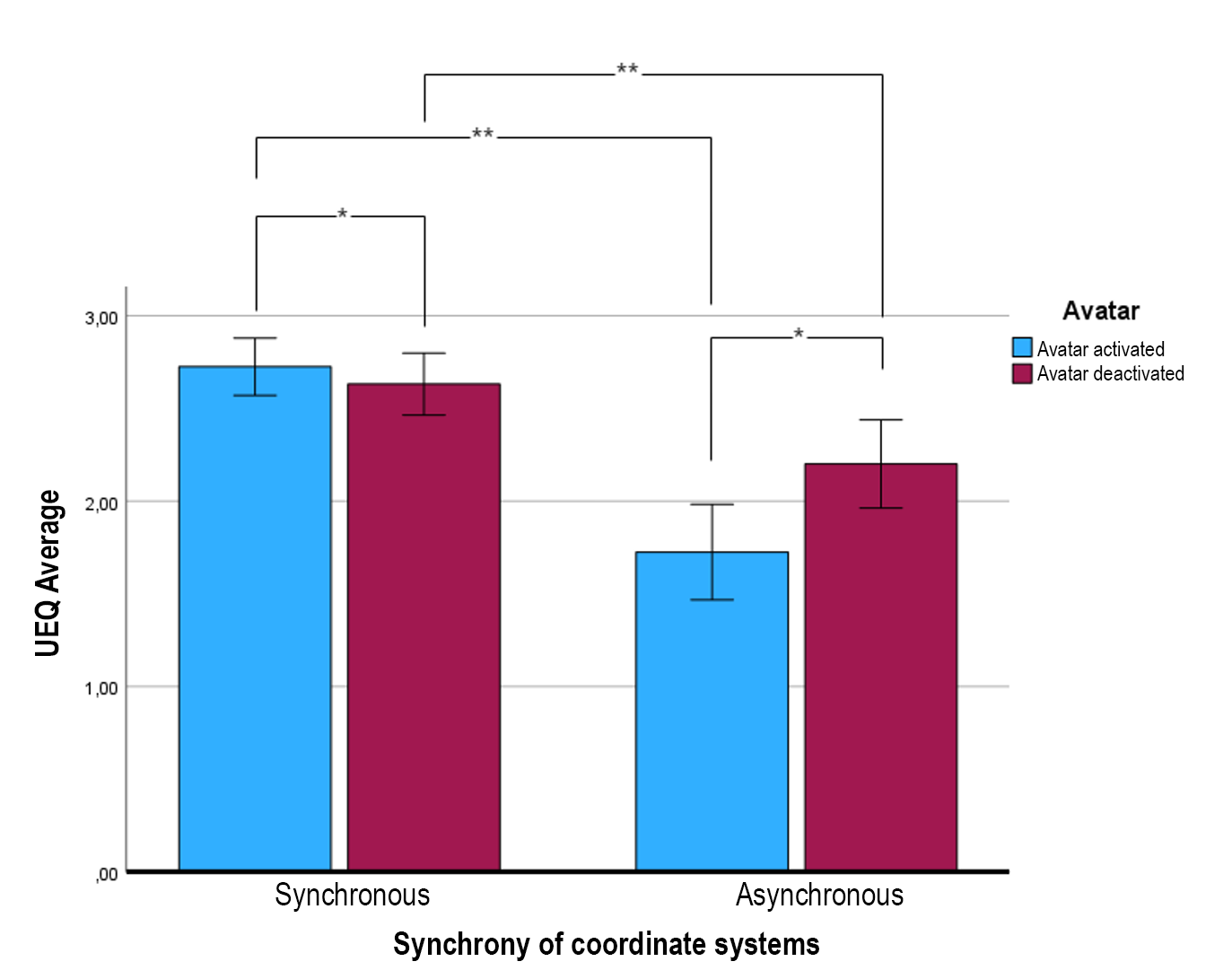}
 \caption{The graph depicts the results from the User Experience Questionnaire (UEQ) for the second experiment. The blue bars represent the average scores where the avatar was activated, and the red bars represent the average scores where the avatar was deactivated. The user experience is perceived significantly better in synchronous conditions compared to asynchronous ones, regardless of whether the avatar is activated or not. Notably, in the synchronous condition, the avatar being activated (blue bar) appears to have a slightly better user experience score than when it is deactivated (red bar). In the asynchronous condition, the avatar's deactivation (red bar) results in a better user experience than when it is activated (blue bar). All the differences here are statistically significant as denoted by the asterisks. \\Significance values: *\textit{p} $<$ .05, **\textit{p} $<$ .01, ***\textit{p} $<$ .001}
 \label{fig:ueq_plot}
\end{figure}
\begin{figure}[tb]
 \centering 
 \includegraphics[width=\columnwidth]{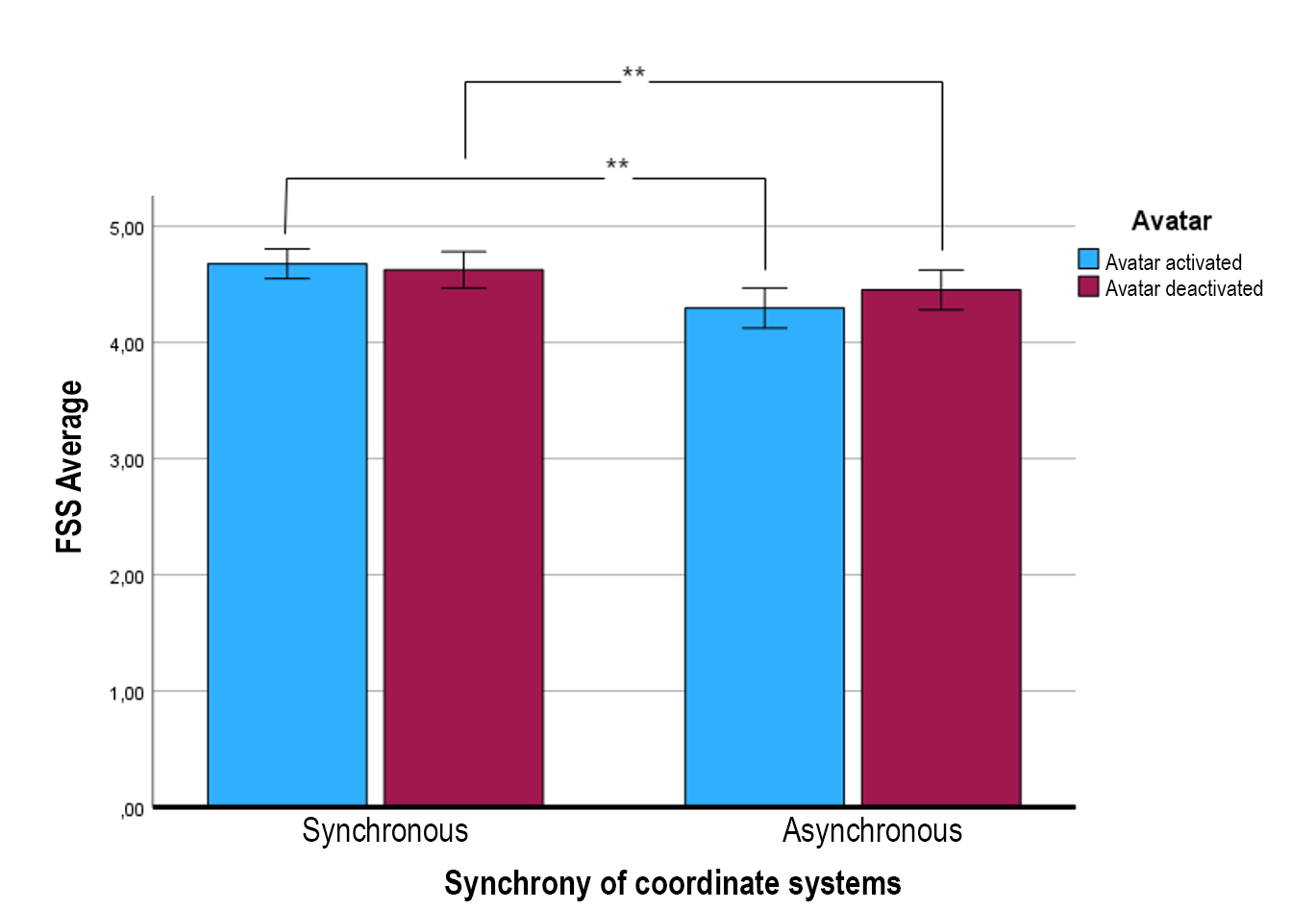}
 \caption{The graph depicts the results from the Flow State Scale (FSS) for the second comparison. The blue bars represent the average scores where the avatar was activated, and the red bars represent the average scores where the avatar was deactivated. For both synchronous and asynchronous conditions, the presence of an avatar (blue) does not show a significant difference in the flow state compared to when the avatar is deactivated (red). 
 The double asterisks (**) denote that the differences between synchronous and asynchronous conditions, both with the avatar activated and deactivated, are statistically significant. \\Significance values: *\textit{p} $<$ .05, **\textit{p} $<$ .01, ***\textit{p} $<$ .001}
 \label{fig:ff_plot}
\end{figure}

\textbf{ATI Analysis.} The average score for \textit{Affinity towards Interactive Technology (ATI)} was $\textit{M} = 3.82$ ($SD= 1.04$). We evaluated the significance of three group variables: gender, age groups, and experience. We adjusted the significance level using the Bonferroni correction \cite{armstrong2014bonferroni} to $\alpha_1 = 0.0166$ to account for multiple comparisons. However, after this adjustment, none of the comparisons reached statistical significance. \\ 
\textbf{UEQ Analysis.} In the context of User Experience \textit{(User Experience Questionnaire - UEQ)}, mean values from two combinations of experimental conditions were examined. The first comparison consists of two conditions that examine the influence of perceptual differences on communication, with a focus on the learnability of the system.:
\begin{itemize}
  \item Identical virtual objects for both users (\textit{Convergence})
  \item One differing virtual object for both users (\textit{Divergence})
\end{itemize}
The mean for the first two conditions (\textit{Convergence \& Divergence}) was $\textit{M} = 2.49$ (${\textit{SD} = 0.62}$). In the first condition (\textit{Convergence}) it was $\textit{M} = 2.61$ ($\textit{SD} = 0.48$) and in the second condition (\textit{Divergence}), $\textit{M} = 2.37$ ($\textit{SD} = 0.72$). The difference was statistically significant, with a two-sided significance level of $p = .027$ (\textit{t}(47) = 2.28). Additionally, an examination was conducted to determine whether gender significantly impacted user experience. For this purpose, five additional analyses were conducted separately, analyzing results for both men and women. These comparisons included Convergence and Divergence for both genders, cross-comparisons of Convergence and Divergence between men and women, and a total cross-comparison. Since the six-fold measurement repetition, the significance level was adjusted using the Bonferroni correction to $\alpha_1 = 0.0083$. However, the results of these comparisons were found to be statistically not significant. 
The second comparison consists of four conditions to investigate the influence of synchronicity and the use of an avatar on collaboration and orientation in an AR environment.
The conditions of the second comparison yielded a mean of $\textit{M} = 2.32$, $\textit{SD} = 0.82$. A two-factorial analysis of variance with repeated measures was conducted to analyze the results from the individual conditions. It was found that both synchronicity ($\textit{F(1,47)} = 63.20$, $p < .001$, $\eta_{\textit{p}}^2 = .57$) and the avatar ($\textit{F(1,47)} = 7.12$, $p = .010$, $\eta_{\textit{p}}^2 = .13$) were significantly associated with the user experience regarding the learnability of the AR environment (Figure \ref{fig:ueq_plot}). Bonferroni-corrected pairwise comparisons of estimated means showed that the synchronous environment ($\textit{M} = 2.678$, $\textit{SE} = .06$) was significantly rated higher than the asynchronous one ($\textit{M} = 1.962$, $\textit{SE} = .11$). The assessment of the use of an avatar ($M = 2.42, SE = 0.09$) or its absence (${M = 2.23}, {SE = 0.08}$) depends on the situation. The avatar was rated lower in the synchronous context, while in the asynchronous context, it was rated higher. The interaction of synchronicity and avatar was also significant ($F(1,47) = 13.07, p < .001, \eta_{\textit{p}}^2 = .22$).\\
\textbf{FSS Analysis.} When examining the flow experience using the \textit{Flow State Scale}, similar to the \textit{UEQ}, individual conditions were tested for their significance. Likewise, six measurements related to the participants' gender were conducted in the first comparison of experimental conditions, but no significant results were found. In the second comparison of experimental conditions, a two-factorial analysis of variance with repeated measures was again conducted. It revealed that synchronicity (${F(1,47) = 16.40}, {p < .001}, {\eta_{\textit{p}}^2 = .25}$) was significantly associated with the experience of flow in the AR environment (Figure \ref{fig:ff_plot}). However, the use of an avatar (${F(1,47) = 1.17}, {p = .286}, {\eta_{\textit{p}}^2 = .024}$) could not be significantly linked to the experience of flow. Bonferroni-corrected pairwise comparisons of estimated means showed that the synchronous environment ($M = 2.68, SE = 0.06$) was significantly rated higher than the asynchronous environment (${M = 1.96}, {SE = 0.11}$). 

\textbf{IPQ Analysis.} In addition to \textit{ATI, UEQ}, and \textit{FSS}, the \textit{IPQ (iGroup presence questionnaire (IPQ)} was used to assess the sense of presence in the virtual environment. The calculated mean was \textit{M} = 3.34 (\textit{SD} = 0.72). Since the results here should also be tested for significance considering the three group variables (gender, age groups, and experience), the significance level of $\alpha=0.05$ was lowered to $\alpha_1=0.0166$ using the Bonferroni correction. The results of the comparisons with the corrected significance level were not significant.

Further on, the results of qualitative analysis have been performed. Responses to open questions were methodically grouped into clusters based on shared themes or recurring statements. Clusters were organized using a Miro board, facilitating a systematic examination of the data. Key themes and patterns were discerned from the analysis of the clusters, which are presented below.

\textbf{Results for Communication Changes.} Participants' responses coalesced into clusters, highlighting the following key themes: (i) \textit{Difficulties in Task Handling}: Participants frequently encountered challenges when communicating in situations where everyone did not share the same visual perspective (6.23\% users); (ii) \textit{Negative Experiences/Emotions:} Some participants reported experiencing negative emotions, such as frustration, confusion, or dissatisfaction, during communication under these conditions (8.34\% users); (iii) \textit{Improvement in Communication:} Others noted positive changes, such as enhanced clarity, precision, or seriousness (41.67\% users); (iv) \textit{Misunderstandings:} Several participants initially experienced misunderstandings but managed to resolve them through adjustments or clarifications (8.34\% users); (v) \textit{Empathy:} Empathizing with the perspectives of others was deemed crucial by some respondents for improved communication and to prevent misunderstandings (8.34\% users); (vi) \textit{Conflict:} Instances of conflict or disputes among participants during communication were reported in some responses (4.17\% users); (vii) \textit{Minimal Change:} Some participants barely noticed any significant change in their communication. The predominant theme was positive statements regarding experience and communication change (10.41\% of users). The remaining responses (12.5\% of users) could not be assigned to any previously mentioned clusters.\\

\textbf{Results for Future with AR/VR.} Responses were categorized into clusters, revealing the following primary themes: (i) \textit{Loss of Reality:} Many participants expressed concerns about a potential loss of their sense of reality due to the extensive use of AR/VR technology (22.92\% users); (ii) \textit{Communication Change/Adaptation:} Participants underscored the necessity for adapting and changing their communication methods to accommodate the impacts of AR/VR technology (29.17\% users); (iii) \textit{Negative Attitude:} A subset of participants conveyed a negative attitude toward anticipated changes in communication, citing issues like confusion, misunderstandings, and other negative consequences (10.41\% users); (iv) \textit{No Change:} Some participants did not anticipate significant alterations in communication patterns resulting from the proliferation of AR/VR technology (10.41\% users). The dominant theme was negative statements regarding isolation and loss of sense of belonging or reality. The remaining responses (10.41\% of users) could not be assigned to any previously mentioned clusters.\\

\textbf{Results for Ideal AR System.} Participants' responses were organized into clusters, revealing the following key themes: (i) \textit{Hardware and Software Optimization:} Many participants expressed a desire for enhanced hardware and software components, including smaller, more manageable, and body-integrated devices, as well as improved screens and cameras (25.00\% users); (ii) \textit{Productivity Aspect:} Some participants viewed the AR system as a valuable tool for daily life and as a means of enhancing productivity (14.59\% users); (iii) \textit{Gaming:} Several responses highlighted the potential for using AR for gaming and entertainment purposes (10.41\% users); (iv) \textit{Vacation/Foreign World:} Participants also considered using an AR system to explore new places or immerse themselves in virtual vacations or fantasy worlds (25.00\% users); (v) \textit{Anti-AR Movement:} A minority of participants fundamentally rejected AR systems. The dominant theme was leisure, fun, and entertainment options (10.41\% of users). The remaining responses (12.5\% of users) could not be assigned to any previously mentioned clusters.\\

\textbf{Observations from Comparisons.} In addition to the data collected during the comparisons, observations were made regarding the behavior and communication of the participants.\\
\textit{Comparison 1.} In the \textit{``Convergence"} condition, participants displayed a positive and engaged demeanor, effectively collaborating to complete tasks. They found the task novel and exciting, leading to successful teamwork.
In contrast, in the \textit{``Divergence"} condition, participants' moods underwent a notable shift, resulting in raised questions and disagreements among them. Mutual questioning was a recurring response in this scenario.\\
\textit{Comparison 2.} Participants employed various forms of navigation to solve tasks, with reactions to avatars spanning from amusement to brief statements. Many participants perceived the avatars' appearance as unclear and abstract, carrying limited significance. The asynchronous environment posed a challenge for all participants, leading to stuttering communication and instilling doubts about their abilities and judgments and those of their navigating partners. Criticism of navigating partners was frequently observed.

\section{Discussion}
Based on the hypothesis for the first comparison (\textit{``Convergence"} and \textit{``Divergence"}), it was assumed that the discovery of differences would lead to increased uncertainty among participants, which, in turn, would result in enhanced communication and cross-checking. The results of the study, along with observations, confirmed this assumption. A significant increase in participant communication was observed, and interactions became more precise. Additionally, heightened scrutiny of other virtual objects was noted. This implies that in the future, divergences in shared AR experiences could be further investigated as a means to increase attention and communication. 

Furthermore, it was surprising that disputes arose among participants in the \textit{``Divergence"} condition. This probably happened due to misunderstandings and disagreements stemming from divergent perspectives. This result highlights the importance of proactive communication and awareness of potential conflicts in divergent shared environments. A practical implication of this is that if an AR application aims to promote different perspectives and enable divergent views, it is crucial to alert participants in advance. This can be done through clear instructions, training, or prompts within the application. By providing this advance notice, participants are made aware that divergent perspectives exist and conflicts may arise. This allows them to proactively deal with these differences, avoid misunderstandings, and ultimately enhance the effectiveness of collaboration. Moreover, since divergent multi-user AR environments are often challenging for users and require effective communication, they could be utilized in practice for team communication training. 

For the second comparison, which examined the synchronicity of user orientations and the use of an avatar, the assumption was that the biggest challenges in task completion would occur in the asynchronous environment without an avatar. It was expected that including the avatar would mitigate the effects of asynchrony. According to the hypothesis, tasks in the \textit{``Sync-w/o-A"} and \textit{``Sync-w-A"} conditions should be completed smoothly and quickly. In the \textit{``ASync-w-A"} environment, somewhat slower task completion was anticipated due to increased complexity but still faster than in the \textit{``ASync-w/o-A"} environment. The results of the study and observations confirmed the hypothesis. Synchronicity was found to have a crucial impact on task completion while using the avatar showed a less significant influence.

Overall, the results of this study provide valuable insights for the design and optimization of shared AR applications.

\section{Conclusion}
The study's findings reveal significant theoretical implications for communication, interaction, and collaboration in VR/AR environments. It examined how varying object perceptions influence communication dynamics, finding no significant impact on interaction. However, qualitative research highlighted the influence of participant conduct and character on collaboration. Furthermore, the study investigated avatars and positional synchrony in VR/AR environments, finding that synchronization is crucial for effective and seamless collaboration, while the use of avatars had no significant impact on interaction. These findings enhance our understanding of communication and user interactions in VR/AR research. Beyond theoretical advancements, they also offer valuable practical insights, guiding the practical application and development of future collaborative AR/VR environments, such as informing participants in advance about divergent perspectives and utilizing divergent multi-user AR environments for team communication training.

\subsection{Limitations and Future Work}
While the current study provides valuable insights, several considerations can enhance the robustness and scope of future research. One notable limitation is the relatively modest sample size. Conducting experiments with a more extensive and diverse group of participants would enhance the precision of the findings and contribute to stronger external validity. 
Another limitation is related to the usage of a specific subset of the questionnaires. Employing the entire questionnaire and introducing a randomized order of experimental conditions would offer a more complete overview of the questionnaire's dimensions and mitigate potential biases in participant responses. 

Looking forward, future research endeavors could focus on addressing these limitations. Efforts to increase sample diversity by conducting experiments with a more diverse and extensive participant pool would capture a broader range of perspectives and experiences.
While the current study focused on in-door experiences, the impact of outdoor activities, such as outdoor AR navigation tasks \cite{voigt2023arnavigation}, could be a promising future research area.
Moreover, extending the study to accommodate up to 30 users interacting simultaneously could offer a more comprehensive understanding of the prototype's dynamics and user experiences in varied social contexts.

\bibliographystyle{IEEEtran}
\bibliography{main}

\begin{thebibliography}{10}
\providecommand{\url}[1]{#1}
\csname url@samestyle\endcsname
\providecommand{\newblock}{\relax}
\providecommand{\bibinfo}[2]{#2}
\providecommand{\BIBentrySTDinterwordspacing}{\spaceskip=0pt\relax}
\providecommand{\BIBentryALTinterwordstretchfactor}{4}
\providecommand{\BIBentryALTinterwordspacing}{\spaceskip=\fontdimen2\font plus
\BIBentryALTinterwordstretchfactor\fontdimen3\font minus \fontdimen4\font\relax}
\providecommand{\BIBforeignlanguage}[2]{{%
\expandafter\ifx\csname l@#1\endcsname\relax
\typeout{** WARNING: IEEEtran.bst: No hyphenation pattern has been}%
\typeout{** loaded for the language `#1'. Using the pattern for}%
\typeout{** the default language instead.}%
\else
\language=\csname l@#1\endcsname
\fi
#2}}
\providecommand{\BIBdecl}{\relax}
\BIBdecl

\bibitem{Azuma1997-wd}
R.~T. Azuma, ``\BIBforeignlanguage{en}{A survey of augmented reality},'' \emph{\BIBforeignlanguage{en}{Presence}}, vol.~6, no.~4, pp. 355--385, Aug. 1997.

\bibitem{Billinghurst2022}
\BIBentryALTinterwordspacing
M.~Billinghurst and H.~Kato, ``Collaborative augmented reality,'' \emph{Commun. ACM}, vol.~45, no.~7, p. 64–70, jul 2002. [Online]. Available: \url{https://doi.org/10.1145/514236.514265}
\BIBentrySTDinterwordspacing

\bibitem{Youjeong2012}
Y.~Kim and S.~Sundar, ``\BIBforeignlanguage{English (US)}{Visualizing ideal self vs. actual self through avatars: Impact on preventive health outcomes},'' \emph{\BIBforeignlanguage{English (US)}{Computers in Human Behavior}}, vol.~28, no.~4, pp. 1356--1364, Jul. 2012.

\bibitem{bowman2007}
D.~Bowman and R.~McMahan, ``Virtual reality: How much immersion is enough?'' \emph{Computer}, vol.~40, pp. 36 -- 43, 08 2007.

\bibitem{7836453}
A.~Irlitti, R.~T. Smith, S.~Von~Itzstein, M.~Billinghurst, and B.~H. Thomas, ``Challenges for asynchronous collaboration in augmented reality,'' in \emph{2016 IEEE International Symposium on Mixed and Augmented Reality (ISMAR-Adjunct)}, 2016, pp. 31--35.

\bibitem{10.3389/frvir.2021.578080}
\BIBentryALTinterwordspacing
M.~Billinghurst, ``Grand challenges for augmented reality,'' \emph{Frontiers in Virtual Reality}, vol.~2, 2021. [Online]. Available: \url{https://www.frontiersin.org/articles/10.3389/frvir.2021.578080}
\BIBentrySTDinterwordspacing

\bibitem{mti6090075}
\BIBentryALTinterwordspacing
M.~Z. Iqbal, E.~Mangina, and A.~G. Campbell, ``Current challenges and future research directions in augmented reality for education,'' \emph{Multimodal Technologies and Interaction}, vol.~6, no.~9, 2022. [Online]. Available: \url{https://www.mdpi.com/2414-4088/6/9/75}
\BIBentrySTDinterwordspacing

\bibitem{Weng2021-ae}
H.~Y. Weng, J.~L. Feldman, L.~Leggio, V.~Napadow, J.~Park, and C.~J. Price, ``\BIBforeignlanguage{en}{Interventions and manipulations of interoception},'' \emph{\BIBforeignlanguage{en}{Trends Neurosci.}}, vol.~44, no.~1, pp. 52--62, Jan. 2021.

\bibitem{Pidel2020CollaborationIV}
\BIBentryALTinterwordspacing
C.~Pidel and P.~Ackermann, ``Collaboration in virtual and augmented reality: A systematic overview,'' in \emph{International Conference on Augmented and Virtual Reality}, 2020. [Online]. Available: \url{https://api.semanticscholar.org/CorpusID:221380739}
\BIBentrySTDinterwordspacing

\bibitem{Szalavari-1998-Stu}
\BIBentryALTinterwordspacing
D.~Schmalstieg, A.~Fuhrmann, Z.~Szalav\'{a}ri, and M.~Gervautz, ``Studierstube - an environment for collaboration in augmented reality,'' 1998, virtual Reality: Research, Development \& Applications, (3): pp. 37-48, 1998. [Online]. Available: \url{https://www.cg.tuwien.ac.at/research/publications/1998/Szalavari-1998-Stu/}
\BIBentrySTDinterwordspacing

\bibitem{regenbrecht2021measuring}
H.~Regenbrecht and T.~Schubert, ``Measuring presence in augmented reality environments: Design and a first test of a questionnaire,'' 2021.

\bibitem{inbook}
X.~Wang and R.~Wang, \emph{Co-presence in Mixed Reality-Mediated Collaborative Design Space}, 01 2011, pp. 51--64.

\bibitem{osmers2021}
\BIBentryALTinterwordspacing
N.~Osmers, M.~Prilla, O.~Blunk, G.~George~Brown, M.~Jan\ss{}en, and N.~Kahrl, ``The role of social presence for cooperation in augmented reality on head mounted devices: A literature review,'' in \emph{Proceedings of the 2021 CHI Conference on Human Factors in Computing Systems}, ser. CHI '21.\hskip 1em plus 0.5em minus 0.4em\relax New York, NY, USA: Association for Computing Machinery, 2021. [Online]. Available: \url{https://doi.org/10.1145/3411764.3445633}
\BIBentrySTDinterwordspacing

\bibitem{voigt2020influence}
J.-N. Voigt-Antons, T.~Kojic, D.~Ali, and S.~M{\"o}ller, ``Influence of hand tracking as a way of interaction in virtual reality on user experience,'' in \emph{2020 Twelfth International Conference on Quality of Multimedia Experience (QoMEX)}.\hskip 1em plus 0.5em minus 0.4em\relax IEEE, 2020, pp. 1--4.

\bibitem{shisha2021}
\BIBentryALTinterwordspacing
A.~H. Hoppe, F.~van~de Camp, and R.~Stiefelhagen, ``Shisha: Enabling shared perspective with face-to-face collaboration using redirected avatars in virtual reality,'' \emph{Proc. ACM Hum.-Comput. Interact.}, vol.~4, no. CSCW3, jan 2021. [Online]. Available: \url{https://doi.org/10.1145/3432950}
\BIBentrySTDinterwordspacing

\bibitem{8797719}
B.~Yoon, H.-i. Kim, G.~A. Lee, M.~Billinghurst, and W.~Woo, ``The effect of avatar appearance on social presence in an augmented reality remote collaboration,'' in \emph{2019 IEEE Conference on Virtual Reality and 3D User Interfaces (VR)}, 2019, pp. 547--556.

\bibitem{ati2017}
T.~Franke, C.~Attig, and D.~Wessel, ``Assessing affinity for technology interaction – the affinity for technology interaction (ati) scale. scale description – english and german scale version,'' 07 2017.

\bibitem{ueq2008}
B.~Laugwitz, T.~Held, and M.~Schrepp, ``Construction and evaluation of a user experience questionnaire,'' vol. 5298, 11 2008, pp. 63--76.

\bibitem{Csikszentmihalyi_undated-fs}
M.~Csikszentmihalyi, ``Flow and the psychology of discovery and invention,'' \url{http://www.mkc.ac.in/pdf/study-material/psychology/2ndSem/UNIT-4-flow-and-creativty-AG.pdf}, accessed: 2023-10-4.

\bibitem{Csikszentmihalyi1992-fh}
M.~Csikszentmihalyi and K.~Rathunde, ``\BIBforeignlanguage{en}{The measurement of flow in everyday life: toward a theory of emergent motivation},'' \emph{\BIBforeignlanguage{en}{Nebr. Symp. Motiv.}}, vol.~40, pp. 57--97, 1992.

\bibitem{Jackson1996-yj}
S.~A. Jackson and H.~W. Marsh, ``\BIBforeignlanguage{en}{Development and validation of a scale to measure optimal experience: The flow state scale},'' \emph{\BIBforeignlanguage{en}{Journal of Sport and Exercise Psychology}}, vol.~18, no.~1, pp. 17--35, Mar. 1996.

\bibitem{ipq2003}
\BIBentryALTinterwordspacing
T.~W. Schubert, ``The sense of presence in virtual environments:,'' \emph{Zeitschrift für Medienpsychologie}, vol.~15, no.~2, pp. 69--71, 2003. [Online]. Available: \url{https://doi.org/10.1026//1617-6383.15.2.69}
\BIBentrySTDinterwordspacing

\bibitem{armstrong2014bonferroni}
\BIBentryALTinterwordspacing
R.~A. Armstrong, ``When to use the bonferroni correction,'' \emph{Ophthalmic and Physiological Optics}, vol.~34, no.~5, pp. 502--508, 2014. [Online]. Available: \url{https://onlinelibrary.wiley.com/doi/abs/10.1111/opo.12131}
\BIBentrySTDinterwordspacing

\bibitem{voigt2023arnavigation}
J.-N. Voigt-Antons, Z.~Sun, M.~Vergari, N.~Ashrafi, F.~Vona, and T.~Kojic, ``Impact of spatial auditory navigation on user experience during augmented outdoor navigation tasks,'' in \emph{2023 Annual congress of the Deutsche Gesellschaft für Akustik (DAGA)}.\hskip 1em plus 0.5em minus 0.4em\relax DEGA Akustik, 2023, pp. 1--4.

\end{thebibliography}

\end{document}